# Field Dependent Conductivity and Threshold Switching in Amorphous Chalcogenides – Modeling and Simulations of Ovonic Threshold Switches and Phase Change Memory Devices

Jake Scoggin, Helena Silva, *Senior Member, IEEE*, and Ali Gokirmak, *Senior Member, IEEE*

*Abstract* – We model electrical conductivity in metastable amorphous $Ge_2Sb_2Te_5$ using independent contributions from temperature and electric field to simulate phase change memory devices and Ovonic threshold switches. 3D, 2D-rotational, and 2D finite element simulations of pillar cells capture threshold switching and show filamentary conduction in the on-state. The model can be tuned to capture switching fields from ~5 to 40 MV/m at room temperature using the temperature dependent electrical conductivity measured for metastable amorphous GST; lower and higher fields are obtainable using different temperature dependent electrical conductivities. We use a 2D fixed out-of-plane-depth simulation to simulate an Ovonic threshold switch in series with a $Ge_2Sb_2Te_5$ phase change memory cell to emulate a crossbar memory element. The simulation reproduces the pre-switching current and voltage characteristics found experimentally for the switch + memory cell, isolated switch, and isolated memory cell.

*Index Terms*—Phase change memory, amorphous semiconductors, finite element analysis

## I. Introduction

PHASE change memory (PCM) is a non-volatile memory that stores information as the conductive crystalline or resistive amorphous phase of a material. The crystalline-to-amorphous phase transition is controllable and reversible, with PCM attaining $10^3$x faster write times and $10^4$x better endurance than flash memory [2]. PCM is CMOS back-end-of-line compatible, allowing memory integration on-chip with CMOS circuitry to eliminate latency from off-chip memory access [3]. PCM can be implemented as a crossbar array, allowing high device density ($4F^2$) in multiple memory layers and efficient neuromorphic computing [4]. Crossbars consist of perpendicular word and bit lines with memory elements sandwiched between these lines at the cross-points (Fig. 1). Each word or bit line is connected to $V_{dd}$ or ground through a transistor, and cross-points can be randomly accessed by activating their corresponding word and bit lines. Non-selected devices can form undesirable current sneak paths between selected and non-selected lines; hence access devices with non-linear current-voltage (*I-V*) characteristics, high $I_{on}/I_{off}$ ratios ($I_{on}/I_{off}$ ~$10^6$ for a 1000x1000 device array), and high drive capabilities to write ($I_{write}$ ~$MA/cm^2$) are needed at each cross-point [5]. Ovonic threshold switches (OTS) made from amorphous chalcogenides are one such access device.

Amorphous chalcogenides are highly resistive under low electric fields and exhibit "threshold switching" to a highly conductive on-state at a threshold voltage ($V_{th}$). Once switched, these materials remain in the on-state while a minimum holding current or voltage ($I_{hold}$ or $V_{hold}$) is maintained (Fig. 2) [6].

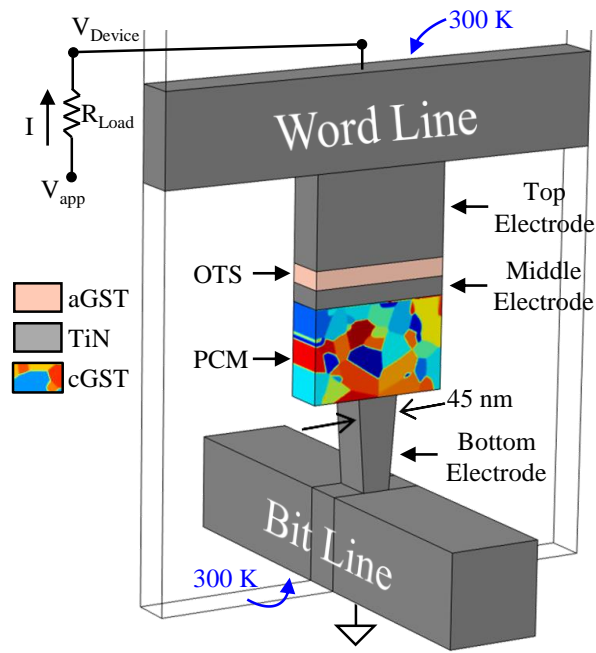

**Fig. 1:** A schematic illustration of a cross-point cell with an Ovonic threshold switch in series with a phase change memory element. The shown cell structure is used for 2D analysis to compare modeling results to the experimental results in [1].

Submitted on XX/XX/XXXX. This work was supported by AFOSR MURI under Award No. FA9550-14-1-0351.

The authors are with the Department of Electrical and Computer Engineering, University of Connecticut, Storrs, CT 06269 USA (email: jacob.scoggin@uconn.edu; helena.silva@uconn.edu; ali.gokirmak@uconn.edu).

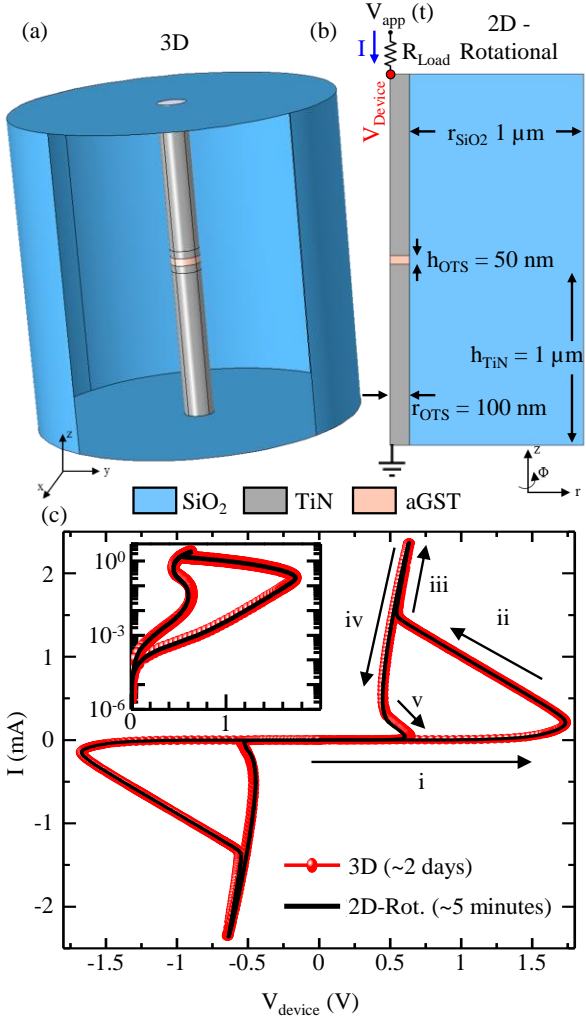

**Fig. 2:** (a) 3D and (b) 2D-rotational OTS geometries used in this work. $T = 300$ K is used as the initial condition and as the boundary condition at the top and bottom of the TiN contacts. (c) When a 3 V / 60 ns triangular pulse is applied at $V_{app}$, the device (i) is highly resistive until $V_{switch} \sim 1.75$ V, (ii) switches on in ~10 ps, (iii - iv) remains on until the voltage and current drop below $V_{hold} \sim 0.4$ V and $I_{hold} \sim 0.25$ mA, and (v) returns to the high resistance off-state. Symmetric switching behavior is observed when a negative ramped pulse is applied. The 2D-rotational and 3D simulations give similar results, as expected for rotationally symmetric filamentary switching. Inset in (c) is the first quadrant in logarithmic scale.

Crystallization dynamics of these materials determine whether they are more suitable for an OTS or a PCM. PCM materials include various stoichiometries of Ag-In-Sb-Te and Ge-Sb-Te, with typical crystallization times on the order of 10 ns [7]. OTS materials remain in their amorphous phase during normal operation and are often characterized by the number of switching cycles they withstand before failure (through, e.g., material damage or crystallization): > 600 for $GeTe_6$ [8], > $10^8$ for AsTeGeSiN [9], and unknown for As-doped Se-Ge-Si, a material used in a commercial OTS+PCM crossbar array [10].

There is still debate on the mechanism(s) underlying threshold switching despite many studies investigating this phenomenon [1], [6], [19], [11]–[18]. $V_{th}$ scales linearly with device thickness, suggesting a field-based mechanism at a threshold $E_{th}$ [15]. Theoretical arguments and the presence of crystalline filaments in failed devices suggest that on-state conduction is filamentary [12], [15].

Current-field ($I$-$\vec{E}$) measurements on amorphous chalcogenides typically show an ohmic regime at low $\vec{E}$, an intermediate regime where $\ln(I) \propto \vec{E}$ or $\vec{E}^{\frac{1}{2}}$, and a high field regime where $I$ increases at a super-exponential rate with $\vec{E}$ [20]. Reference [20] reviews conduction mechanisms in amorphous materials and shows that multiple mechanisms can fit measured data through the tuning of parameters which are otherwise difficult to validate (e.g. trap-to-trap distance, effective carrier mass, and carrier mobility). Models for these mechanisms have been proposed which define carrier concentrations ($n$) and mobilities ($\mu$) as functions of $\vec{E}$ and temperature ($T$) such that all three $I$-$\vec{E}$ regimes are captured with an electrical conductivity $\sigma = qn\mu$, but such techniques are computationally expensive in addition to relying on multiple unknown fitting parameters. Reference [13] proposes a field-based switching model where carrier concentrations rapidly increase once trap states near the Fermi band are filled and fits the model to amorphous (a-) $Ge_2Sb_2Te_5$ (GST) measurements. References [18], [19] propose a field-assisted thermal model based on multiple trap barrier lowering and fit the model to a-GeTe and doped a-GST measurements. References [14], [21] ascribe switching to crystalline filaments which form under high $\vec{E}$ and fit the model to a-GST using relaxation oscillations. They suggest that these filaments become unstable at low $\vec{E}$ in OTS materials but remain stable in PCM materials.

Here, we model conductivity in a-GST as the sum of $T$ and $\vec{E}$ dependent terms ($\sigma_a = \sigma_T + \sigma_E$). This model does not require a computationally expensive evaluation of the (density of states × Fermi function) integral at every $T$, $\vec{E}$ combination where conductivity is needed; hence, it is appropriate for transient finite element simulations with dynamic $T$ and $\vec{E}$. While this model trades accuracy for ease of computation, simulations show that it (i) can be tuned to fit a wide range of switching fields, (ii) captures the appropriate changes in threshold switching as we systematically vary ambient conditions, geometries, and the rise and fall times of applied pulses, and (iii) can reproduce the behavior of a series PCM+OTS device when used with a finite element phase change model [22]–[25].

## II. COMPUTATIONAL MODEL

We use a-GST material parameters as in [23], [25] to simulate an OTS material. Of particular interest to this work is $\sigma_a$, which we model as in [26] and cap at $\sigma_{max} = \sigma_T(930\ K)$ (Fig. 3):

$$\sigma_a(T, \vec{E}) = \min(\sigma_T(T) + \sigma_E(\vec{E}),\ \sigma_{max}), \quad (1)$$

which is equivalent to assuming that free carriers are excited to

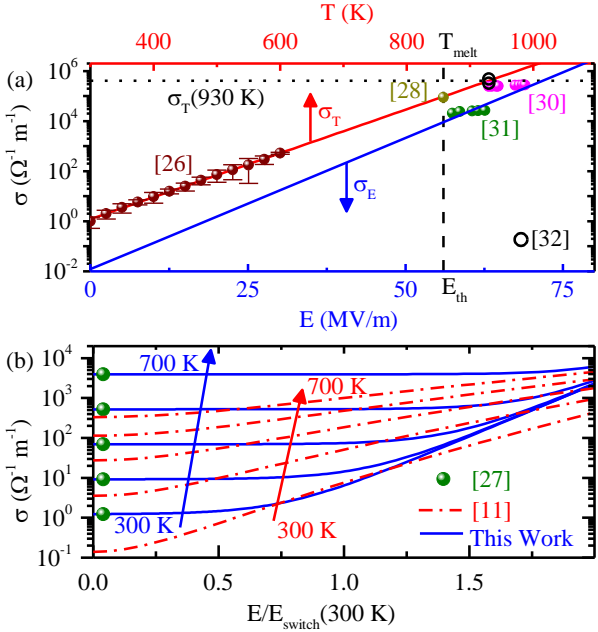

**Fig. 3:** (a) $T$ and $\vec{E}$ dependent contributions to electrical conductivity in (1); temperature is uncertain for $T > T_{melt}$. (b) Conductivity-Field behavior using the model in this work (metastable a-GST) and the model in [19] (drifted, doped a-GST) at various temperatures. We use $E_{switch}$(300 K) = 25.01 MV / m in this work and $E_{switch}$(300 K) = 155 MV/m for the curves in [19].

a band edge via independent thermal and electrical processes:

$$\sigma_a(T, \vec{E}) = \min(\, q(n_T + n_E)\mu \,,\, \sigma_{max} \,), \quad (2)$$

where $n_T$ and $n_E$ are carriers excited via thermal or electrical processes and $\mu$ is the free carrier mobility.

We fit $\sigma_T$ to low-$\vec{E}$ measurements of metastable a-GST wires [27] and molten GST thin films [28], as described in [29] (Fig. 3a). Measurements of liquid GST show a semiconductor-to-metal transition near 930 K [30], with $\sigma$ becoming practically independent of $T$. We therefore limit $\sigma_a(T, \vec{E})$ to $\sigma_T$(930 K) = $4.1\times10^5$ [$\Omega^{-1}$ m$^{-1}$], which is in line with the highest conductivities measured in molten GST [30]–[32](Fig. 3a).

$\sigma_E$ is assumed to be an exponential which contributes 1% of $\sigma_T$(300 K) at zero field and 10% of $\sigma_T$($T_{melt}$) at $E_{th}$:

$$\sigma_E(\vec{E}) = \frac{\sigma_T(300\,K)}{100} \exp(|\vec{E}| \cdot C_1) \quad (3)$$

where $C_1 = 2.42\times10^{-7}$ m/V is chosen such that $\sigma_E(E_{th}) = \sigma_T(T_{melt}) \times 10\%$. We use $E_{th}$ = 56 MV/m, the breakdown field measured in as-deposited a-GST [33], and $T_{melt}$ = 858 K [28] (Fig. 3a).

We include $\sigma$-$\vec{E}$ curves at various temperatures calculated using the models in this work (for metastable a-GST) and the models in [19] (for drifted, doped a-GST) for comparison (Fig. 3b). $\sigma_T$ dominates at low fields, while $\sigma_E$ begins to dominate at higher and higher fields with increasing T.

We couple heat transfer and current continuity physics to simulate transient device operation, including temperature dependent thermoelectric effects [34]:

$$d_m c_p \frac{dT}{dt} - \nabla \cdot (k\nabla T) = -\nabla V \cdot J - \nabla \cdot (JST) + q_H \quad (4)$$

$$\nabla \cdot J = \nabla \cdot (-\sigma \nabla V - \sigma S \nabla T) = 0 \quad (5)$$

where $d_m$ is mass density, $c_p$ is specific heat, $t$ is time, $k$ is thermal conductivity, $V$ is electric potential, $J$ is current density, $S$ is the Seebeck coefficient, and $q_H$ accounts for the latent heat of phase change. We use temperature dependent parameters for a-GST, crystalline GST (c-GST), TiN, and SiO$_2$ as in [23], [25].

We model phase change as

$$\frac{d\overrightarrow{CD}}{dt} = \overrightarrow{Nucleation} + \overrightarrow{Growth} + \overrightarrow{Melt}, \quad (6)$$

where $\overrightarrow{CD}$ is a 2-vector whose magnitude ($CD$) corresponds to phase ($CD$ = 0 or 1 for the amorphous or crystalline phase) and whose orientation ($\theta_{CD}$) corresponds to grain orientation ($\tan(\theta_{CD}) = CD_2/CD_1$), capturing nucleation, growth, and grain boundary melting [23], [25]. We solve for (6) in PCM devices but not OTS devices, which we assume do not crystallize in our simulations.

### III. SIMULATIONS

We first simulate switching in 3D and 2D-rotational geometries by applying a 3V / 60 ns triangular pulse (Fig. 2a,b). Results show filamentary switching with current confined to the central portion of the device (Fig. 4a-j), with practically identical *I-V* characteristics for 3D and 2D-rotational simulations (Fig. 2c). 3D simulations show some instability of the filament location (slightly off-center in Fig. 4d,i) and filament migration over time.

We next evaluate the impact of $\sigma_E$ by simulating switching with $\sigma_E$ = 0 (Fig. 5a) and compare the results with $\sigma_E$ defined as in (3) (Fig. 5b). $V_{switch}$ and $I_{switch}$ are the values at which $V_{Device}$ begins to decrease. Threshold switching occurs even with $\sigma_E$ = 0 due to thermal runaway. However, defining $\sigma_E$ as in (3) gives a switching field ($E_{switch} = V_{switch} / h_{OTS}$) that is smaller and less dependent on $h_{OTS}$ (Fig. 5c). Some $h_{OTS}$ dependence is still observed due to changing thermal conditions.

Reference [33] reports switching fields from 8.1 MV/m (as-deposited a-Ge$_{15}$Sb$_{85}$) to 94 MV/m (as-deposited 4 nm thick a-Sb). We examine the tunability of our model by varying $E_{th}$ in (3) from 5.6 to 560 MV/m (Fig. 6). Results show $E_{switch}$ varying from 5 to 42.5 MV/m. $E_{switch}$ = 25.01 MV/m when $E_{th}$ = 56 MV/m, similar to the $E_{switch}$ = 28.75 MV/m measured in [13] for melt-quenched a-GST. $\sigma_E$ becomes negligible compared to $\sigma_T$ when $E_{th}$ > 200 MV/m even for high fields: the $\sigma_T$ used in this work precludes $E_{switch}$ > 42.5 MV/m; a reduced $\sigma_T$ is required for higher switching fields. $I_{switch}$ decreases by ~100x as $V_{switch}$ increases, resulting in a decrease in switching power ($P_{switch}$) from ~100 to 20 μW (Fig. 6c).

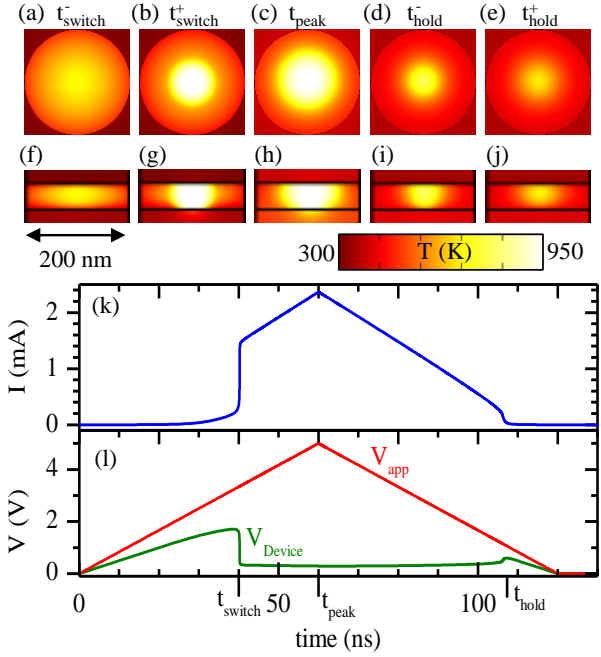

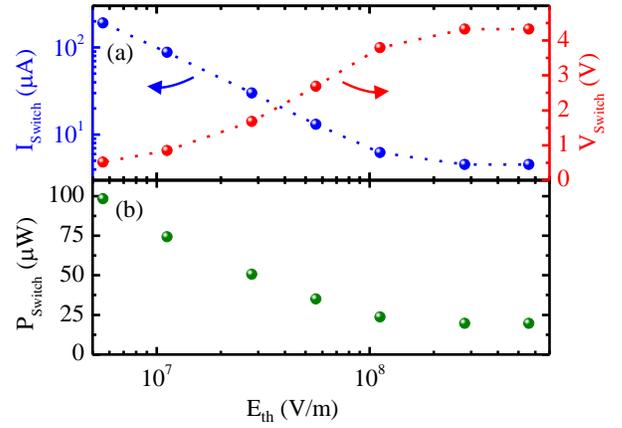

**Fig. 4:** (a-e) x-y and (f-j) x-z temperature cut planes while switching the 3D OTS in Fig. 2a illustrate filamentary on-state conduction. (j) Current and (k) device voltage transients resulting from the applied $V_{app}$ used to generate the I-V in Fig. 2c. Superscripts "-" and "+" refer to the time steps (1 ns increments) before and after the subscripted event. (Fig. 2a: $R_{Load}$ = 1 kΩ, $h_{OTS}$ = 50 nm, $r_{OTS}$ = 100 nm, $V_{app}$ = 5 V / 60 ns triangular pulse).

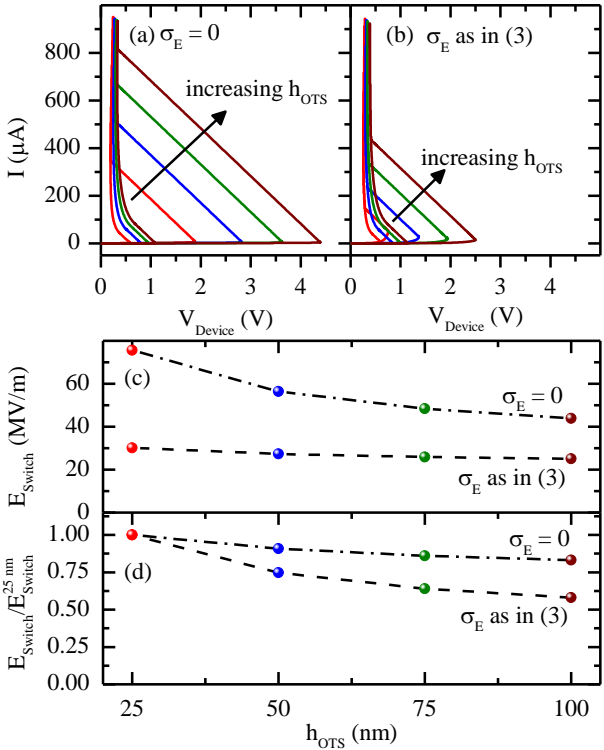

**Fig. 5:** The switching voltage increases with $h_{OTS}$ both (a) without and (b) with field dependent conductivity. Including field dependent conductivity reduces the switching field's (c) magnitude and (d) sensitivity to $h_{OTS}$. (Fig. 2b: $R_{Load}$ = 5 kΩ, $h_{OTS}$ = 25 to 100 nm, $r_{OTS}$ = 100 nm, $V_{app}$ = 5 V / 5 s triangular pulse).

**Fig. 6:** (a) The switching current (voltage) decreases (increases) and (c) the switching power decreases with increasing $E_{th}$ in (3). The device switches thermally before the field contribution becomes significant for $E_{th}$ > 1×10$^8$ V / m. As a result, further increases to $E_{th}$ result in the same switching characteristics. (Fig. 2b: $R_{Load}$ = 5 kΩ, $h_{OTS}$ = 100 nm, $r_{OTS}$ = 100 nm, $V_{app}$ = 5 V / 5 s triangular pulse).

$V_{switch}$ has been shown to decrease with increasing ambient temperature ($T_{ambient}$), while the temperature behavior of $I_{switch}$ and $P_{switch}$ are less clear [19]. We simulate switching while varying $T_{ambient}$ (the initial temperature and the fixed top and bottom TiN boundary temperatures, Fig. 2b) from 300 to 400 K (Fig. 7). $V_{switch}$ decreases as expected (Fig. 7a). $I_{switch}$ at first decreases and then increases with increasing $T_{ambient}$, while $P_{switch}$ monotonously decreases in the $T_{ambient}$ range simulated but is beginning to flatten with increasing $T$ by 400 K.

Next, we systematically vary $r_{OTS}$, $h_{OTS}$, and the rise time ($\tau_{rise}$) of $V_{app}$ in the geometry shown in Fig. 2b (Fig. 8). Results agree with expected OTS behavior: $V_{switch}$ approximately doubles as $h_{OTS}$ doubles [6], $V_{switch}$ decreases with increasing $\tau_{rise}$, approaching a minimum value [16], $I_{hold}$ and $V_{hold}$ are only weakly dependent on $h_{OTS}$ [15], and switching characteristics are only weakly dependent on $r_{OTS}$ due to filamentary conduction in the on-state.

Finally, we simulate an OTS and PCM in series (OTS+PCM, Fig. 1) based on the devices fabricated and characterized in [1]. We use a 2D, 45 nm fixed out-of-plane-depth simulation instead of a 2D-rotational simulation to more appropriately model phase change dynamics in the PCM with (6). We use a 500 nm depth in the bit line to account for its large thermal mass (Fig. 1), set $T_{ambient}$ = 300 K as the initial and fixed TiN boundary temperatures, and reset the device with a 5 V / 5 ns square pulse (1 ns rise and fall times) at $V_{app}$ followed by 1 μs for thermalization (starting at Fig. 1 and ending at Fig. 9a). We then sweep $V_{app}$ from 0 to 2.5 V over 50 ns to characterize the OTS + reset PCM. We also simulate an isolated OTS and isolated reset PCM in the same way by replacing the PCM or OTS, respectively, with TiN (Fig. 9c,d). We plot the I-V characteristics before switching, dividing currents and voltages by the isolated OTS $I_{switch}$ and $V_{switch}$ values in order to compare

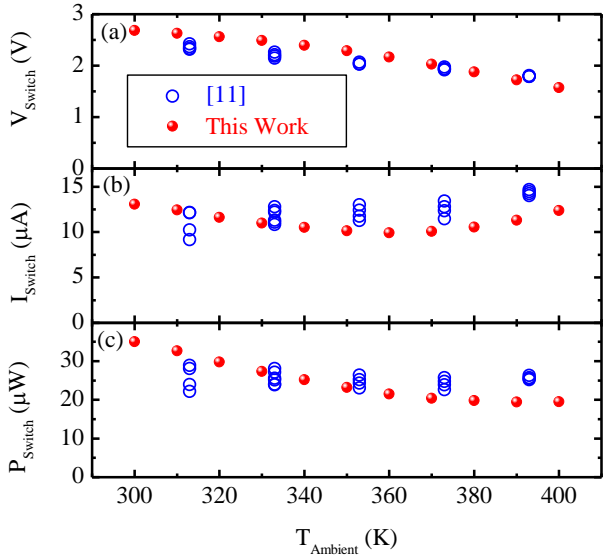

**Fig. 7:** The (a) voltage, (b) current, and (c) power required to switch as the ambient temperature changes. $V_{switch}$ decreases monotonically, but $I_{switch}$ and $P_{switch}$ have more complex relationships with $T_{ambient}$. (Fig. 2b: $R_{Load}$ = 5 kΩ, $h_{OTS} = r_{OTS}$ = 100 nm, $V_{app}$ = 5 V / 5 s triangular pulse).

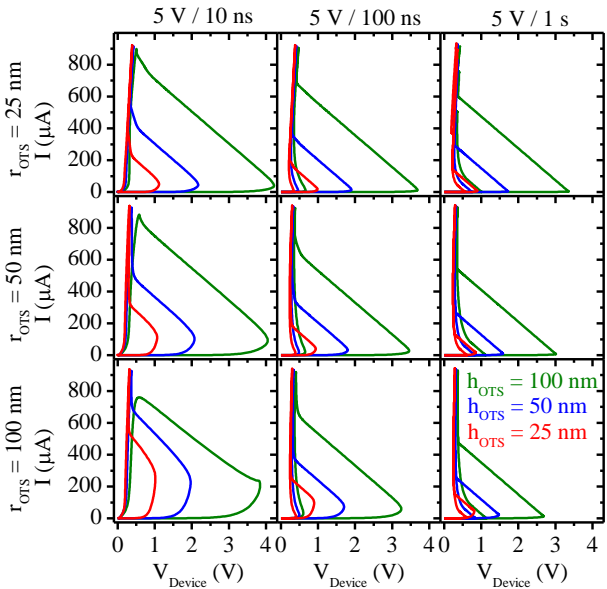

**Fig. 8:** Switching characteristics of OTS devices with varying radii, ramp times, and heights. The simulated holding currents and voltages are weakly dependent on $h_{OTS}$, but the switching voltage ~doubles as as $h_{OTS}$ doubles. (Geometry in Fig. 2b).

our results to those presented in [1] (Fig. 9e). The results are similar, with the OTS limiting the current in the reset PCM until $V_{Device} / V_{switch}^{OTS} > 2$ (Fig. 9d). This limits the current during read in a reset cell while allowing high current in a set cell, creating a large read margin. The smaller scaled currents for the PCM and OTS+PCM in our simulation could be due to a difference in the amorphous volume in the reset PCM; the fixed out-of-plane depth in our simulations, which cannot capture filaments smaller than 45 nm in depth and may thus overestimate $I_{switch}$ in the OTS; or parameter differences

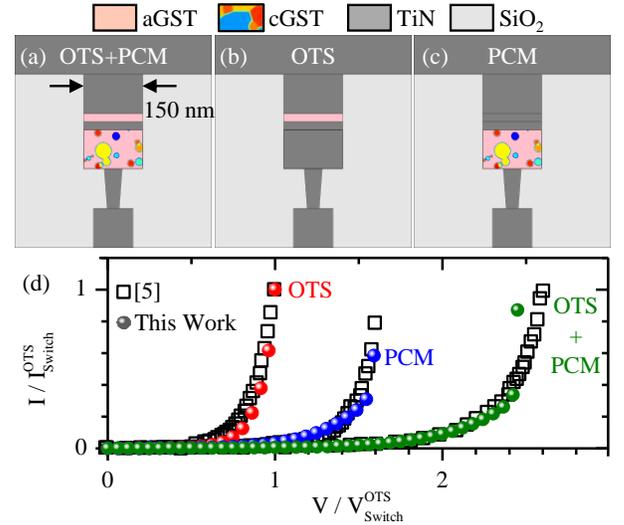

**Fig. 9:** (a) Reset OTS+PCM, (b) isolated OTS, and (c) isolated reset PCM. (d) Pre-switching I-V characteristics scaled by $V_{switch}$ and $I_{switch}$ in the OTS using the model in this work (spheres) and experimental data extracted from [1] (squares). I-V characteristics from this work show similar switching voltages but lower scaled switching currents. The schematic of the simulation setup is shown in Fig. 1. The OTS+PCM reset animation for this simulation is available in supplementary material.

between aGST and the (unreported) OTS material used in [1].

## IV. CONCLUSION

The coupling of thermal and electric field contributions to electrical conductivity in amorphous semiconductors is complex, as evidenced by the large number of physical models proposed to explain the same characteristics. Our modeling results show that threshold switching and 'snap-back' observed in OTS and PCM devices can be explained through electro-thermal phenomena giving rise to thermal run-away and filamentary conduction and can be modeled efficiently with a finite element framework.

## V. ACKNOWLDEGEMENTS

The authors would like to thank Ilya Karpov of Intel Corporation, Martin Salinga of RWTH Aachen University, Abu Sebastian of IBM Zurich, and Geoffrey Burr of IBM Almaden for valuable discussions.

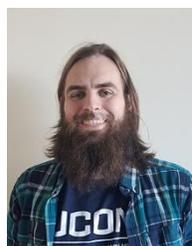

**Jake Scoggin** received his B.S. degree in Nanosystems Engineering from Louisiana Tech University in 2012 and his M.S. degree in Electrical Engineering from Louisiana Tech University in 2014. He is a PhD student in Electrical & Computer Engineering at the University of Connecticut since 2014.

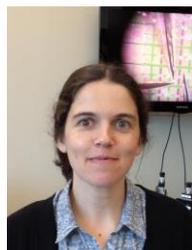

**Helena Silva** received her B.S. degree in Engineering Physics from the University of Lisbon, Portugal in 1998 and her M.S. and Ph.D. degrees in Applied Physics from Cornell University in 2002 and 2005. She is currently an Associate Professor of Electrical and Computer Engineering at the University of Connecticut. Her research focuses on electronics devices for logic, memory, and energy conversion.

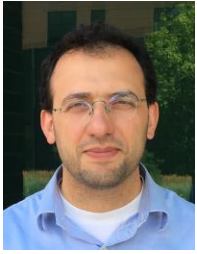

**Ali Gokirmak** received his B.S. degrees in Electrical Engineering and Physics from University of Maryland at College Park in 1998 and received his Ph.D. in Electrical and Computer Engineering from Cornell University in 2005. He joined the Electrical & Computer Engineering Department of the University of Connecticut in 2006, where he is currently an associate professor .